\documentclass[prb,preprint,onecolumn,showpacs,preprintnumbers,amsmath,amssymb,superscriptaddress,floatfix]{revtex4-2}
\usepackage[T1]{fontenc}
\usepackage[ansinew]{inputenc}
\usepackage{graphics}
\usepackage{dcolumn}
\usepackage{bm}
\usepackage{amsthm}
\usepackage{amsmath}
\usepackage{amssymb}
\usepackage{diagbox}
\usepackage{hyperref}
\usepackage{url}
\usepackage{xcolor}

\begin{document}

\title{Magnetic neutron scattering from spherical nanoparticles with N\'{e}el surface anisotropy: Atomistic simulations}
 
\author{Michael P.\ Adams}\email[Electronic address: ]{michael.adams@uni.lu}
\affiliation{Department of Physics and Materials Science, University of Luxembourg, 162A~avenue de la Faiencerie, L-1511~Luxembourg, Grand Duchy of Luxembourg}

\author{Andreas Michels}\email[Electronic address: ]{andreas.michels@uni.lu}
\affiliation{Department of Physics and Materials Science, University of Luxembourg, 162A~avenue de la Faiencerie, L-1511~Luxembourg, Grand Duchy of Luxembourg}  
 
\author{Hamid Kachkachi}\email[Electronic address: ]{hamid.kachkachi@univ-perp.fr}
\affiliation{Universit\'{e} de Perpignan via Domitia, Laboratoire PROMES CNRS UPR8521, Rambla de la Thermodynamique, Tecnosud, F-66100~Perpignan, France} 
 

\begin{abstract}
We consider a dilute ensemble of randomly-oriented noninteracting spherical nanomagnets and investigate its magnetization structure and ensuing neutron-scattering response by numerically solving the Landau-Lifshitz equation. Taking into account the isotropic exchange interaction, an external magnetic field, a uniaxial magnetic anisotropy for the particle core, and in particular the N\'{e}el surface anisotropy, we compute the magnetic small-angle neutron scattering cross section and pair-distance distribution function from the obtained equilibrium spin structures. The numerical results are compared to the well-known analytical expressions for uniformly magnetized particles and provide guidance to the experimentalist. Moreover, the effect of a particle-size distribution function is modeled.
\end{abstract}

\date{\today}

\maketitle

 
\section{Introduction}

Magnetic nanoparticles are the subject of intense worldwide research efforts which are partly motivated by potential applications in \textit{e.g.}\ medicine, biology, and nanotechnology (see, \textit{e.g.}\ Refs.~\cite{lakbenderdisch2021,diebold2010applications,de2008applications,baetke2015applications,stark2015industrial,han2019applications,BATLLE2022} and references therein). In the majority of studies, the internal spin structure of the nanoparticles is neglected and assumed to be uniform (so-called macro- or superspin model). While this is probably justified in many application-oriented approaches in which an overall understanding is sufficient, it is of interest, at least from the standpoint of fundamental science, to elucidate the effect of a nonuniform spin structure on a certain physical property. 

Scattering techniques, in particular employing x-rays and neutrons, have proven to be very powerful in this endeavor, since they provide statistically-averaged information on a large number of scattering particles. For instance, using Monte Carlo simulations of a discrete atomistic spin model, K\"ohler~et~al.~\cite{koehlerjac2021} have numerically studied the influence of antiphase boundaries in iron oxide nanoparticles on their spin structure. These authors used the Debye scattering equation to relate the internal spin disorder to the broadening of certain x-ray Bragg peaks. Vivas~et~al.~\cite{laura2020} carried out micromagnetic continuum calculations of the spin structure of defect-free iron nanoparticles and related a vortex-type magnetization configuration to certain signatures in the magnetic neutron scattering cross section and correlation function.

Magnetic small-angle neutron scattering (SANS) is a powerful technique for investigating spin structures on the mesoscopic length scale ($\sim 1$$-$$100 \, \mathrm{nm}$) and inside the volume of magnetic materials~\cite{rmp2019,michelsbook}. Recent SANS studies of magnetic nanoparticles, in particular employing spin-polarized neutrons, unanimously demonstrate that their spin textures are highly complex and exhibit a variety of nonuniform, canted, or core-shell-type configurations (see, \textit{e.g.}\ Refs.~\cite{disch2012,kryckaprl2014,ijiri2014,guenther2014,maurer2014,dennis2015,grutter2017,oberdick2018,krycka2019,benderapl2019,bersweiler2019,zakutna2020,dirkreview2022} and references therein). The magnetic SANS data analysis largely relies on structural form-factor-models for the cross section, borrowed from nuclear SANS, which do not properly account for the existing spin inhomogeneity inside a magnetic nanoparticle. Progress in magnetic SANS theory~\cite{michels2013,michels2014jmmm,mettus2015,erokhin2015,metmi2015,metmi2016,michelsPRB2016,michelsdmi2019,mistonov2019,metlov2022} strongly suggests that for the analysis of experimental magnetic SANS data, the spatial nanometer scale variation of the orientation and magnitude of the magnetization vector field must be taken into account, going beyond the macrospin-based models that assume a \textit{uniform} magnetization.

In this paper, we employ \textit{atomistic} simulations using the Landau-Lifshitz equation (LLE) to investigate the role of the N\'{e}el surface anisotropy in magnetic nanoparticles and its effect on the magnetic SANS cross section and correlation function. We take into account the isotropic exchange interaction, an external magnetic field, a magnetocrystalline anisotropy for the core of the nanoparticles and N\'{e}el's anisotropy for spins on the surface. Moreover, the influence of a particle-size distribution function on the magnetic SANS cross section and pair-distance correlation function is studied. The numerical results reveal marked differences with the superspin model and provide guidance for the experimentalist to identify nonuniform spin structures inside magnetic nanoparticles. We also refer to our analytical study of the problem~\cite{adamsjacana2022}, which is restricted to a linear approximation in the magnetization deviation.

The paper is organized as follows:~In Section~\ref{sec2}, we provide information on the atomistic simulations using the LLE. In Section~\ref{sec3}, we display the expressions for the magnetic SANS cross section and for the pair-distance distribution function. The results of the numerical calculations are discussed in Section~\ref{sec4}, with Section~\ref{sec4p1} focusing on the effect of the N\'{e}el surface anisotropy and Section~\ref{sec4p2} discussing the influence of a lognormal particle-size distribution on the SANS observables. Section~\ref{sec5} summarizes the main findings of this study and provides an outlook on future challenges.

\section{Details of the atomistic SANS modeling using the Landau-Lifshitz equation}
\label{sec2}

\begin{figure*}[tb!]
\centering
\resizebox{1.0\columnwidth}{!}{\includegraphics{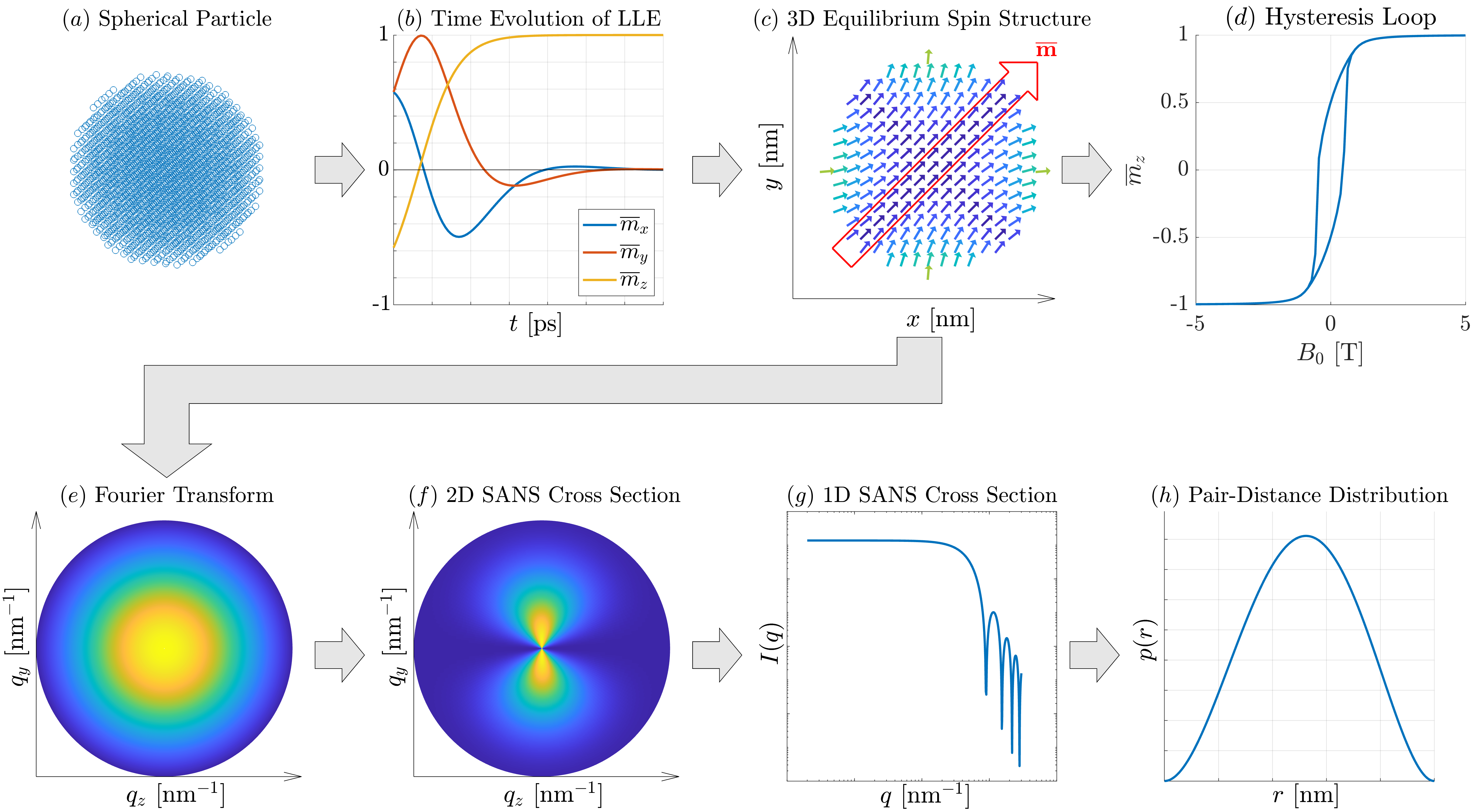}}
\caption{Flow chart explaining the atomistic SANS simulation procedure. (\textit{a})~A spherical nanoparticle is cut from a simple cubic grid with $N \times N \times N$ atoms. (\textit{b})~Time evolution of the Cartesian magnetization components obtained by solving the Landau-Lifshitz equation. (\textit{c})~Computed equilibrium spin structure of a spherical nanoparticle at remanence (cut through the center of the particle). (\textit{d})~Hysteresis loop of an ensemble of randomly-oriented nanoparticles. (\textit{e})~Computed Fourier transform and (\textit{f})~two-dimensional magnetic SANS cross section $d\Sigma_M / d\Omega$. (\textit{g})~Azimuthally-averaged magnetic SANS cross section $I(q)$ and (\textit{h})~pair-distance distribution function $p(r)$.}
\label{fig1}
\end{figure*}

Fig.~\ref{fig1} schematically depicts the adopted procedure to generate and calculate the spin structure, and to obtain the ensuing magnetic SANS cross section and correlation function. This flowchart-type representation will be discussed in more detail in the following.

A spherical many-spin nanomagnet is viewed as a crystallite consisting of $\mathcal{N}$ atomic magnetic moments $\boldsymbol{\mu}_i = \mu_a \mathbf{m}_i$, where $\mu_{a}$ denotes the magnitude of the atomic magnetic moment and $\mathbf{m}_i$ is a unit vector specifying its orientation. We assume the spins to `sit' on a simple cubic lattice, so that $\mu_a = M_s a^3$, where $M_s$ is the saturation magnetization of the material and $a$ is the lattice constant. The spherical shape of the nanomagnet is cut from a simple cubic regular grid [Fig.~\ref{fig1}(\textit{a})], and its radius $R$ is defined as $R = \frac{N-1}{2} a$, where the integer $N$ is the number of atoms on the side of the cubic grid. The magnetic state of the nanomagnet is investigated with the help of the atomistic approach based on the following Hamiltonian~\cite{dimwys94prb,kodber99prb,kacgar01physa300,kacgar01epjb,igllab01prb,kacdim02prb,kacgar05springer,kazantsevaetal08prb}:
\begin{align}
\mathcal{H} &= \mathcal{H}_{{\mathrm{EX}}} + \mathcal{H}_{\mathrm{Z}} + \mathcal{H}_{\mathrm{A}} \\
&= -\frac{1}{2} J \sum_{i,j \in{\mathrm{n.n.}}} \mathbf{m}_{i} \cdot \mathbf{m}_{j} - \mu_{a} \mathbf{B}_0 \cdot \sum_{i=1}^{\mathcal{N}} \mathbf{m}_{i} + \sum_{i=1}^{\mathcal{N}}\mathcal{H}_{\mathrm{A},i} ,
\label{eq:Ham-MSP}
\end{align}
where $\mathcal{H}_{{\mathrm{EX}}}$ is the nearest-neighbor (n.n.) exchange energy, with $J > 0$ the exchange parameter, $\mathcal{H}_{{\mathrm{Z}}}$ denotes the Zeeman energy, with $\mathbf{B}_0 = \mu_0 \mathbf{H}_0$ the homogeneous externally applied magnetic field, and $\mathcal{H}_{{\mathrm{A}}}$ represents the magnetic anisotropy energy. For the core spins, we assume the anisotropy to be of uniaxial symmetry, while for surface spins we adopt the model proposed by N\'{e}el~\cite{nee54jpr}. $\mathcal{H}_{\mathrm{A},i}$ can then be expressed as follows:
\begin{equation}
\mathcal{H}_{\mathrm{A},i} = \begin{cases}
-K_{c} \left(\mathbf{m}_{i} \cdot \mathbf{e}_{\mathrm{A}} \right)^{2}, & i \in{\mathrm{core}}
\\ \\ + \frac{1}{2}K_{s}{\displaystyle \sum_{j\in{\mathrm{n.n.}}}}\left(\mathbf{m}_{i}\cdot\mathbf{u}_{ij}\right)^{2}, & i\in\mathrm{surface} ,
\end{cases}
\label{eq:HamUA-NSA}
\end{equation}
where $K_c > 0$ and $K_s > 0$ denote, respectively, the core and surface anisotropy constants, $\mathbf{e}_{\mathrm{A}}$ is a unit vector along the core anisotropy easy direction, and $\mathbf{u}_{ij} = (\mathbf{r}_i - \mathbf{r}_j) / \|\mathbf{r}_i - \mathbf{r}_j \|$ is a unit vector connecting the nearest-neighbor spins $i$ and $j$.

The magnetodipolar interaction has been ignored in our simulations. This is motivated by the numerical complexity of this energy term, in particular for atomistic simulations (here for a $10 \, \mathrm{nm}$~diameter particle the number of spins is $\mathcal{N} = 11633$), and by the expectation that it is of minor relevance for smaller-sized nanomagnets~\cite{koehlerjac2021,hertel2021}.

The dynamics of each individual magnetic moment $\mathbf{m}_{i}$ is described by the Landau-Lifshitz  equation (LLE)~\cite{berkovinkronparkinhandbook07}:
\begin{equation}
\frac{d{\mathbf{m}}_i}{dt} = - \gamma \, \mathbf{m}_{i} \times \mathbf{B}_{i}^{\mathrm{eff}} - \alpha \, \mathbf{m}_{i} \times (\mathbf{m}_{i} \times \mathbf{B}_{i}^{\mathrm{eff}}) ,
\label{eq:LLLE}
\end{equation}
where $\gamma$ is the gyromagnetic ratio, and $\alpha$ denotes the damping constant. The deterministic effective magnetic field acting on the spin $i$ is given by:
\begin{align}
\mathbf{B}_i^{\mathrm{eff}} &= - \frac{1}{\mu_a} \frac{\delta \mathcal{H}}{\delta \mathbf{m}_i}
\nonumber
\\
&= 
\mathbf{B}_0 + \frac{J}{\mu_a} \sum_{j \in \mathrm{n.n.}} \mathbf{m}_{j}
\nonumber
\\
&-
\frac{1}{\mu_a}
\begin{cases}
-2K_c (\mathbf{m}_i \cdot \mathbf{e}_{\mathrm{A}}) \, \mathbf{e}_{\mathrm{A}},
& \;\;\; i \in \text{core},
\\
K_s \sum_{j \in \mathrm{n.n.}} (\mathbf{m}_i \cdot \mathbf{u}_{ij}) \, \mathbf{u}_{ij},
& \;\;\; i \in \text{surface}.
\end{cases}
\end{align}
The LLE is numerically solved by using the explicit Euler-forward-projection method~\cite{bavnas2005numerical}, which consist of two steps. The first step, as seen from equation~\eqref{eq:EulerForward} below, is the simple Euler forward scheme and the second step, as seen from equation~\eqref{eq:EulerForwardProjectionStep} is the projection (or normalization) onto the unit sphere to enforce the constraint $\|\mathbf{m}_i\| = 1$. Since we are interested in the static equilibrium, this first-order method is fully appropriate. In equations~\eqref{eq:EulerForward} and \eqref{eq:EulerForwardProjectionStep}, $k$ is the time iteration index while $i$ refers to the $i$th lattice site:
\begin{align}
\mathbf{m}_{i}^{\mathrm{Euler}} &= \mathbf{m}_{i}^{k} + h_t \, \frac{d{\mathbf{m}}_i^k}{dt}
\label{eq:EulerForward} ,
\\
\mathbf{m}_{i}^{k+1} &= \frac{\mathbf{m}_{i}^{\mathrm{Euler}} }{\|\mathbf{m}_{i}^{\mathrm{Euler}} \|}
\label{eq:EulerForwardProjectionStep} ,
\end{align}
where $h_t$ denotes the time step for the integration procedure. For the termination of the energy minimization, we have employed the following criterion:
\begin{equation}
\label{Micromagn_minproccriterion}
\frac{h_t}{\mathcal{N}} \sqrt{\sum_{i=1}^{\mathcal{N}} \left \| \frac{d{\mathbf{m}}_i^k}{dt} \right \|^2} < 10^{-8} .
\end{equation}
The macroscopic state of the nanomagnet is then described by the following super- or macrospin (representing the net magnetic moment):
\begin{equation}
\overline{\mathbf{m}} = \frac{1}{\mathcal{N}}\sum_{i=1}^{\mathcal{N}}\mathbf{m}_{i} .
\label{eq:Macrospin}
\end{equation}
As an example, we show in Fig.~\ref{fig1}(\textit{b}) the temporal evolution of the Cartesian magnetization components of $\overline{\mathbf{m}}$ and in Fig.~\ref{fig1}(\textit{c}) the numerically-computed equilibrium spin configuration for a spherical nanomagnet at zero applied field, in a plane across its center. It is seen that the spins in the center of the nanoparticle are directed along $\overline{\mathbf{m}}$, while the surface spins exhibit significant misalignment, which is due to the presence of the N\'{e}el surface anisotropy. Note that the $\mathbf{m}_i$ are unit vectors, whereas generally $\|\overline{\mathbf{m}}\| \neq 1$.

In our simulations, we used the following parameters:~atomic magnetic moment $\mu_{a} = 1.577 \times 10^{-23}$~Am$^2$ (corresponding to $1.7 \, \mu_\mathrm{B}$ with $\mu_\mathrm{B}$ the Bohr magneton), lattice constant $a = 0.3554 \, \mathrm{nm}$, $M_s = 351 \, \mathrm{kA/m}$, exchange constant $J = 8.7 \times 10^{-22} \, \mathrm{J/atom}$, core anisotropy constant $K_c = 3 \times 10^{-24} \, \mathrm{J/atom}$, damping constant $\alpha = 3 \times 10^{11}$~(Ts)$^{-1}$, gyromagnetic constant $\gamma = 1.76 \times 10^{11}$~(Ts)$^{-1}$, and an integration time step of $h_t = 5$~fs. The surface anisotropy constant $K_s$ was used as an adjustable parameter.

For the calculation of the magnetic SANS cross section $d\Sigma_M / d\Omega$ [Fig.~\ref{fig1}(\textit{d})], it is necessary to compute the discrete Fourier transform of all the ${\mathbf{m}}_i$ belonging to the spherical nanomagnet [Fig.~\ref{fig1}(\textit{e})]. In Sec.~\ref{sec3}, the expressions for $d\Sigma_M / d\Omega$ are formulated for a continuous magnetization distribution $\mathbf{M}(\mathbf{r})$ and of its Fourier transform $\widetilde{\mathbf{M}}(\mathbf{q})$. These functions are defined as follows:
\begin{align}
\mathbf{M}(\mathbf{r}) &= \frac{1}{(2\pi)^{3/2}} \int \widetilde{\mathbf{M}}(\mathbf{q}) \exp\left(\mathrm{i} \mathbf{q} \cdot \mathbf{r} \right) \; d^3q , \\
\widetilde{\mathbf{M}}(\mathbf{q}) &= \frac{1}{(2\pi)^{3/2}} \int \mathbf{M}(\mathbf{r}) \exp\left(- \mathrm{i} \mathbf{q} \cdot \mathbf{r} \right) \; d^3r .
\end{align}
Using $\boldsymbol{\mu}_{i} = \mu_{a} \mathbf{m}_{i}$, the discrete-space Fourier transform is computed as: 
\begin{align}
\widetilde{\mathbf{M}}(\mathbf{q}) \cong \frac{\mu_a}{(2\pi)^{3/2}} \sum_{i=1}^{\mathcal{N}} \mathbf{m}_i \exp\left(-\mathrm{i} \mathbf{q} \cdot \mathbf{r}_i \right) ,
\label{discreteFT}
\end{align}
where $\mathbf{r}_i$ is the location point of the $i$th spin and $\mathbf{q}$ represents the wave vector (scattering vector). Equation~(\ref{discreteFT}) establishes the relation between the outcome of the simulations, $\mathbf{m}_i$, and the magnetic SANS cross section $d\Sigma_M / d\Omega$. In the standard SANS geometry, the $\mathbf{q}$-space of interest is defined by $\mathbf{q} = q [0, \, \sin \theta,\, \cos\theta]$, which corresponds to the two-dimensional detector plane ($q_x = 0$, see Fig.~\ref{fig2}). The two- and one-dimensional magnetic SANS cross section $d\Sigma_M / d\Omega$ [Fig.~\ref{fig1}(\textit{f}) and (\textit{g})] is then computed according to equation~(\ref{eq:equation1}). A further Fourier transformation yields the pair-distance distribution function [Fig.~\ref{fig1}(\textit{h})].

At each value of the external field, atomistic simulations of the spin structure and of the ensuing magnetic SANS cross section were carried out for $256$ random orientations of the core anisotropy axes $\mathbf{e}_{\mathrm{A}}$ of the particle with respect to the field $\mathbf{B}_0$. More specifically, once the lattice orientation has been randomly selected, the easy-axis orientation of the particle's core and the distribution of the N\'{e}el anisotropy are fixed. The whole system (core plus surface anisotropy) is then randomly rotated relative to $\mathbf{B}_0$. For the generation of the random angles, we used the low-discrepancy Sobol sequence~\cite{sobol}. Therefore, except Fig.~\ref{fig3}, all the data shown in this paper correspond to an ensemble of randomly-oriented particles. The simulations were carried out by starting from a large positive (saturating) field of about $10 \, \mathrm{T}$ and then the field was reduced in steps of typically $30 \, \mathrm{mT}$.

\section{Magnetic SANS cross section and pair-distance distribution function}
\label{sec3}

\begin{figure}[tb!]
\centering
\resizebox{0.70\columnwidth}{!}{\includegraphics{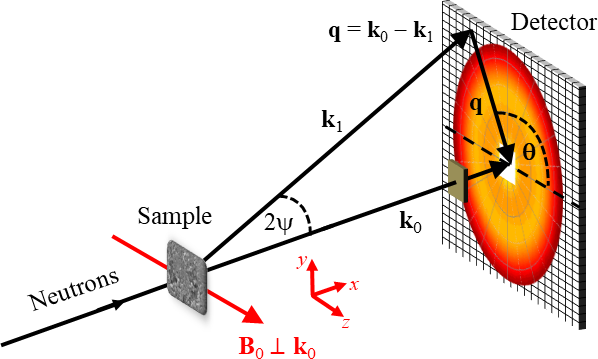}}
\caption{Sketch of the neutron scattering geometry. The applied magnetic field $\mathbf{B}_0 \parallel \mathbf{e}_z$ is perpendicular to the wave vector $\mathbf{k}_0 \parallel \mathbf{e}_x$ of the incident neutron beam ($\mathbf{B}_0 \perp \mathbf{k}_0$). The momentum-transfer or scattering vector $\mathbf{q}$ is defined as the difference between $\mathbf{k}_0$ and $\mathbf{k}_1$, \textit{i.e.}\ $\mathbf{q} = \mathbf{k}_0 - \mathbf{k}_1$. SANS is usually implemented as elastic scattering ($k_0 = k_1 = 2\pi / \lambda$), and the component of $\mathbf{q}$ along the incident neutron beam, here $q_x$, is much smaller than the other two components, so that $\mathbf{q} \cong [0, q_y, q_z] = q [0, \sin\theta, \cos\theta]$. This demonstrates that SANS probes predominantly correlations in the plane perpendicular to the incident beam. For elastic scattering, the magnitude of $\mathbf{q}$ is given by $q = (4\pi / \lambda) \sin(\psi)$, where $\lambda$ denotes the mean wavelength of the neutrons and $2\psi$ is the scattering angle. The angle $\theta = \angle(\mathbf{q}, \mathbf{B}_0)$ is used to describe the angular anisotropy of the recorded scattering pattern on the two-dimensional position-sensitive detector.}
\label{fig2}
\end{figure}

The quantity of interest in experimental SANS studies is the elastic magnetic differential scattering cross section $d \Sigma_M / d \Omega$, which is usually recorded on a two-dimensional position-sensitive detector. For the most commonly used scattering geometry in magnetic SANS experiments, where the applied magnetic field $\mathbf{B}_0 \parallel \mathbf{e}_z$ is perpendicular to the wave vector $\mathbf{k}_0 \parallel \mathbf{e}_x$ of the incident neutrons (see Fig.~\ref{fig2}), $d \Sigma_M / d \Omega$ (for unpolarized neutrons) can be written as~\cite{rmp2019}:
\begin{eqnarray}
\frac{d \Sigma_M}{d \Omega}(\mathbf{q}) = \frac{8 \pi^3}{V} b_H^2 \left( |\widetilde{M}_x|^2 + |\widetilde{M}_y|^2 \cos^2\theta \right. \nonumber \\ \left. + |\widetilde{M}_z|^2 \sin^2\theta - (\widetilde{M}_y \widetilde{M}_z^{\ast} + \widetilde{M}_y^{\ast} \widetilde{M}_z) \sin\theta \cos\theta \right) ,
\label{eq:equation1}
\end{eqnarray}
where $V$ is the scattering volume, $b_H = 2.91 \times 10^8 \, \mathrm{A}^{-1}\mathrm{m}^{-1}$ is the magnetic scattering length in the small-angle regime (the atomic magnetic form factor is approximated by $1$ since we are dealing with forward scattering), $\widetilde{\mathbf{M}}(\mathbf{q}) = [\widetilde{M}_x(\mathbf{q}), \widetilde{M}_y(\mathbf{q}), \widetilde{M}_z(\mathbf{q})]$ represents the Fourier transform of the magnetization vector field $\mathbf{M}(\mathbf{r}) = [M_x(\mathbf{r}), M_y(\mathbf{r}), M_z(\mathbf{r})]$, $\theta$ denotes the angle between $\mathbf{q}$ and $\mathbf{B}_0$, and the asterisk `$*$' stands for the complex-conjugated quantity. Note that in the perpendicular scattering geometry the Fourier components are evaluated in the plane $q_x = 0$ (see Fig.~\ref{fig2}).

The numerically computed magnetic SANS cross sections that are displayed in this paper correspond to the following average:
\begin{eqnarray}
\frac{d \Sigma_M}{d \Omega} = \left\langle\frac{d \Sigma_M}{d \Omega}\right\rangle_{\mathbf{e}_{\mathrm{A}}} = \frac{1}{\mathcal{K}} \sum_{k=1}^{\mathcal{K}} \frac{d \Sigma_{M,k}}{d \Omega},
\label{eq:equation1average}
\end{eqnarray}
where $d \Sigma_{M,k} / d \Omega$ represents (for fixed $K_s$ and $B_0$) the magnetic SANS cross section for a particular core easy-axis orientation $\mathbf{e}_{\mathrm{A}}$ (referred to index `$k$'), and $\mathcal{K}$ denotes the number of random configurations. Equation~(\ref{eq:equation1average}) implies the absence of interparticle interactions.

For a uniformly magnetized spherical particle with its saturation direction parallel to $\mathbf{e}_z$, \textit{i.e.}\ $M_x = M_y = 0$ and $M_z = M_s$, equation~(\ref{eq:equation1}) reduces to:
\begin{eqnarray}
\frac{d \Sigma_M}{d \Omega}(q, \theta) = V_p (\Delta \rho)_{\mathrm{mag}}^2 \, 9 \left( \frac{j_1(qR)}{qR} \right)^2 \sin^2\theta ,
\label{homomagsans2}
\end{eqnarray}
where $V_p = \frac{4\pi}{3} R^3$ is the particle's volume, $(\Delta \rho)_{\mathrm{mag}}^2 = b_H^2 \left( \Delta M \right)^2 = b_H^2 M_s^2$ is the magnetic scattering-length density contrast, and $j_1(qR)$ is the first-order spherical Bessel function. The well-known analytical result for the homogeneous sphere case, equation~(\ref{homomagsans2}), and its correlation function [see equation~(\ref{pvonreq}) below] serve as a reference for comparison with the nonuniform case.

It is often convenient to average the two-dimensional SANS cross section $\frac{d \Sigma_M}{d \Omega}(\mathbf{q}) = \frac{d \Sigma_M}{d \Omega}(q_y, q_z) = \frac{d \Sigma_M}{d \Omega}(q, \theta)$ along certain directions in $\mathbf{q}$-space, \textit{e.g.}\ parallel ($\theta = 0$) or perpendicular ($\theta = \pi/2$) to the applied magnetic field, or even over the full angular $\theta$~range. In the following, we consider the $2\pi$-azimuthally-averaged magnetic SANS cross section
\begin{eqnarray}
\label{aziaverage}
I(q) \equiv \frac{1}{2\pi} \int_0^{2\pi} \frac{d \Sigma_M}{d \Omega}(q,\theta) \, d\theta ,
\end{eqnarray}
which is used to compute the pair-distance distribution function $p(r)$ according to:
\begin{eqnarray}
\label{pvonreqintegral}
p(r) = r^2 \int\limits_0^{\infty} I(q) j_0(qr) q^2 dq,
\end{eqnarray}
where $j_0(qr) = \sin(qr)/(qr)$ is the spherical Bessel function of zero order; $p(r)$ corresponds to the distribution of real-space distances between volume elements inside the particle weighted by the excess scattering-length density distribution; see the reviews by Glatter~\cite{glatterchapter} and by Svergun and Koch~\cite{svergun03} for detailed discussions of the properties of $p(r)$ and for information on how to compute it by indirect Fourier transformation~\cite{bender2017}. For our discrete simulation data, the integrals in equations~(\ref{aziaverage}) and (\ref{pvonreqintegral}) were approximated by the trapezoidal rule. Apart from constant prefactors, the $p(r)$ of the azimuthally-averaged single-particle cross section [equation~(\ref{homomagsans2})], corresponding to a uniform sphere magnetization, is given by (for $r \leq 2R$):
\begin{eqnarray}
\label{pvonreq}
p(r) = r^2 \left( 1 - \frac{3r}{4R} + \frac{r^3}{16 R^3} \right) .
\end{eqnarray}
We also display results for the correlation function $c(r)$, which is related to $p(r)$ by
\begin{eqnarray}
\label{cvonreq}
c(r) = p(r)/r^2 .
\end{eqnarray}
As we will demonstrate in the following, when the particles' spin structure is inhomogeneous, the $d \Sigma_M / d \Omega$ and the corresponding $p(r)$ and $c(r)$ differ significantly from the homogeneous case [equations~(\ref{homomagsans2}) and (\ref{pvonreq})], which serve as a reference. Due to the $r^2$~factor, features in $p(r)$ at medium and large distances are more pronounced than in $c(r)$.

\section{Results and Discussion}
\label{sec4}

\subsection{Effect of the N\'{e}el surface anisotropy}
\label{sec4p1}

\begin{figure}[tb!]
\centering
\resizebox{1.0\columnwidth}{!}{\includegraphics{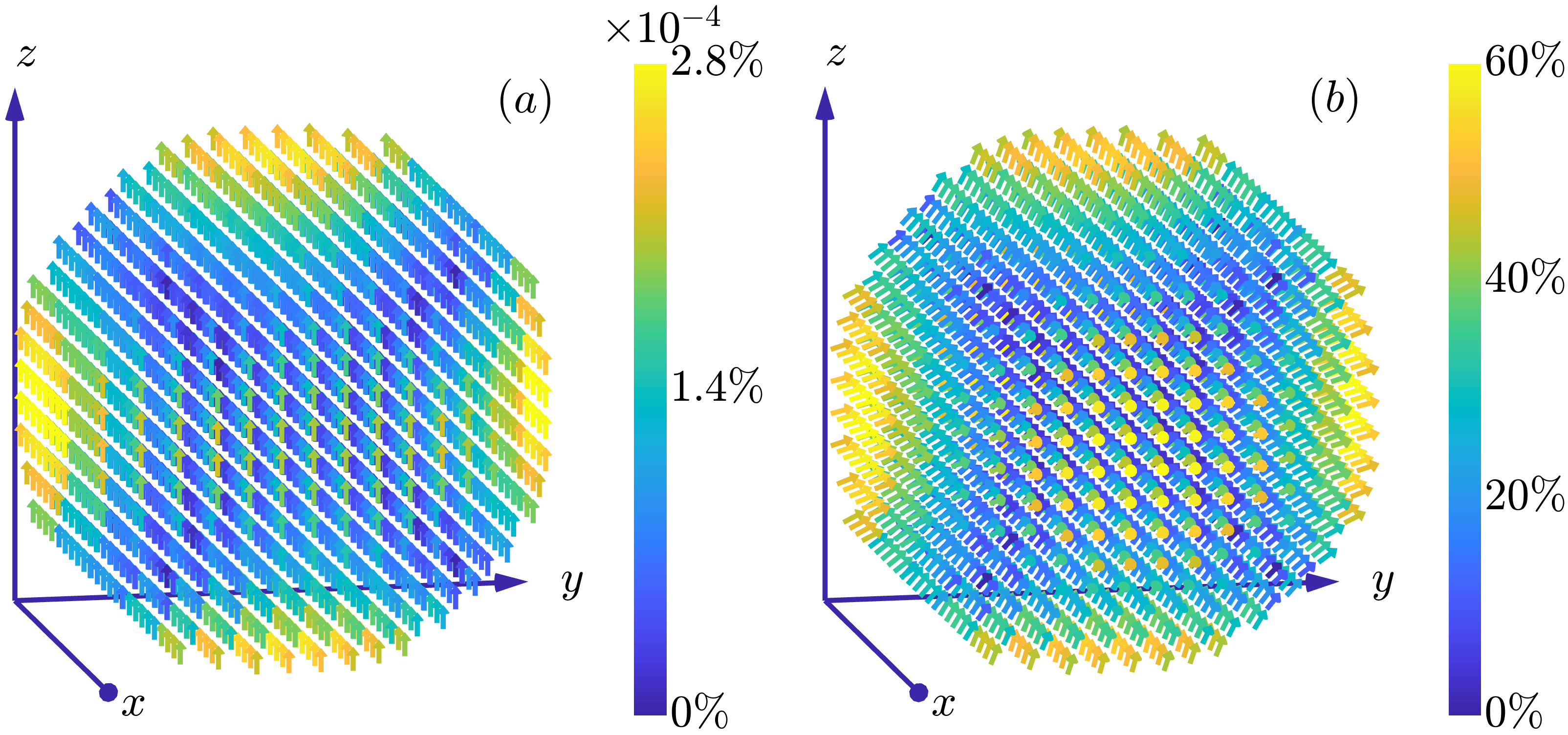}}
\caption{Selected 3D equilibrium spin structures arising from the N\'{e}el surface anisotropy (compare also to Figs.~2 and 3 in the analytical study~\cite{adamsjacana2022}). (\textit{a})~$K_s = 5.22 \times 10^{-23} \, \mathrm{J/atom}$; (\textit{b})~$K_s = 52.2 \times 10^{-23} \, \mathrm{J/atom}$. Further parameters are: core-anisotropy axis $\mathbf{e}_{\mathrm{A}} = [0,0,1]$, core-anisotropy constant $K_c = 3 \times 10^{-24} \, \mathrm{J/atom}$, and external magnetic field $\mathbf{B}_0 = [0,0,150 \, \mathrm{mT}]$. The color code depicts the spin misalignment relative to the average magnetization vector, namely $\delta m_j = \| \mathbf{m}_j - \overline{\mathbf{m}}/\|\overline{\mathbf{m}}\| \|$. At the surface of the nanomagnet the spin deviations are larger compared to the core.}
\label{fig3}
\end{figure} 

\begin{figure}[tb!]
\centering
\resizebox{0.80\columnwidth}{!}{\includegraphics{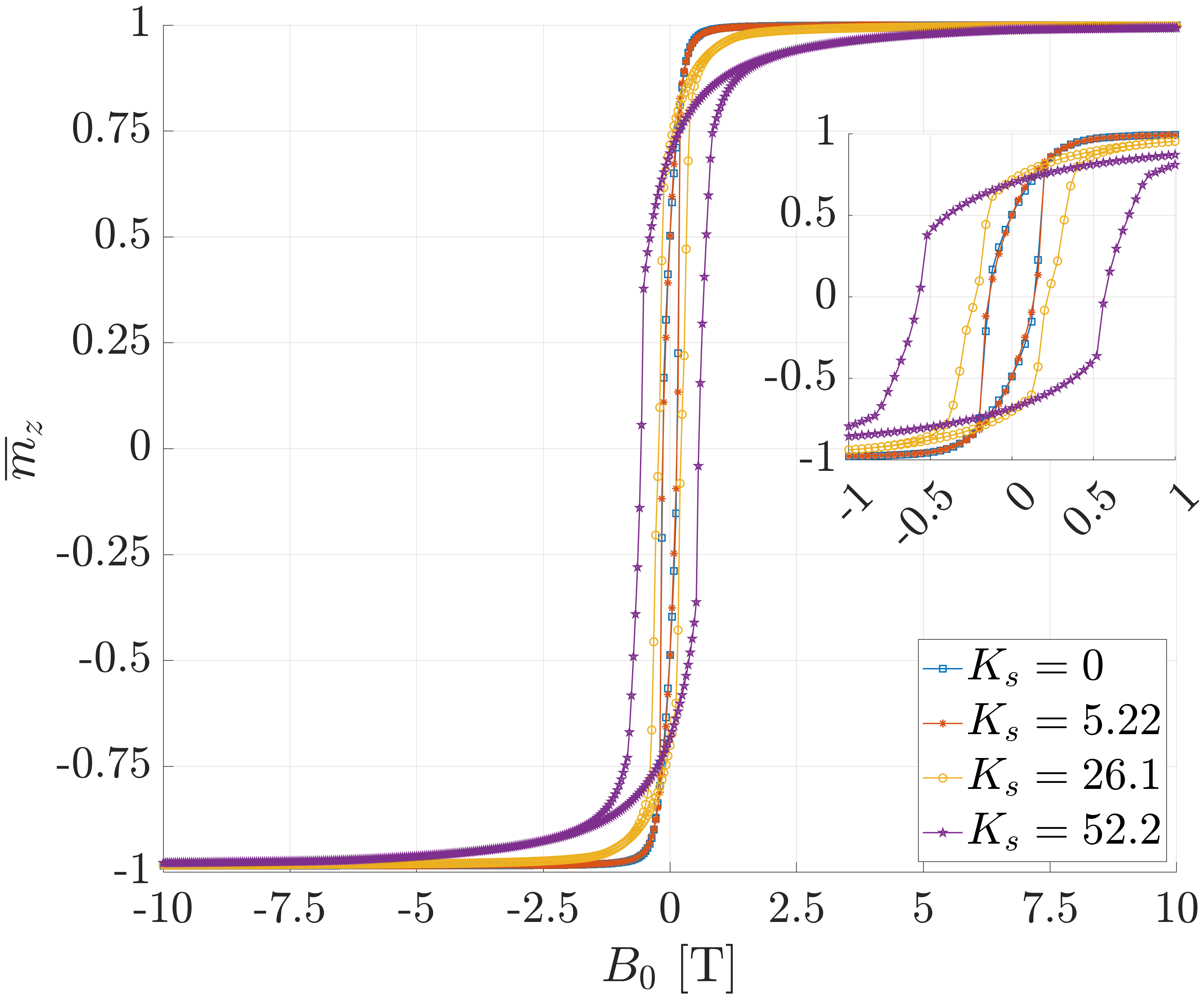}}
\caption{Computed normalized magnetization $\overline{m}_z$ [compare equation~(\ref{eq:Macrospin})] of an ensemble of randomly-oriented $10$-nm-sized spherical nanomagnets for different values of the surface anisotropy constant $K_s$ (in units of $10^{-23} \, \mathrm{J/atom}$, see inset).}
\label{fig4}
\end{figure}

Fig.~\ref{fig3} displays as an example the spin structures of a $5$-nm-sized spherical nanomagnet for the cases of a small and large surface-anisotropy constant $K_s$, and Fig.~\ref{fig4} shows computed hysteresis curves for an ensemble of randomly-oriented $10$-nm-sized nanomagnets. As expected, increasing $K_s$ results, for a given particle size, in a progressive surface spin disorder which propagates into the bulk of the nanomagnet. The effect of an enhanced $K_s$ also becomes visible in the magnetization curves via an increased coercivity $H_c$ and remanence $m_r$. For $K_s = 0$ and dominant exchange, we recover the well-known results from the Stoner-Wohlfarth model~\cite{usov1997}, \textit{i.e.}\ we find a reduced remanence of $m_r = 0.5$ and a coercivity of
\begin{eqnarray}
\label{swhcadapted}
\mu_0 H_c = 0.48 \frac{2K_c}{\mu_a} \frac{\mathcal{N}_c}{\mathcal{N}} = 183 \, \mathrm{mT} ,
\end{eqnarray}
where $\mathcal{N}_c$ denotes the number of atoms belonging to the particle's core. Note that for the case of a strong surface anisotropy [Fig.~\ref{fig3}(\textit{b})], the mean magnetization at remanence deviates strongly from the core anisotropy axis, which is in contrast to the case of weak anisotropy [Fig.~\ref{fig3}(\textit{a})]. This observation is in agreement with the analytical calculations by Garanin and Kachkachi~\cite{garanin2003} who have predicted the emergence of an effective anisotropy of \textit{cubic} symmetry for dominant $K_s$. Therefore, with increasing $K_s$, we observe in Fig.~\ref{fig4} an increase of the remanence.

\begin{figure*}[tb!]
\centering
\resizebox{1.0\columnwidth}{!}{\includegraphics{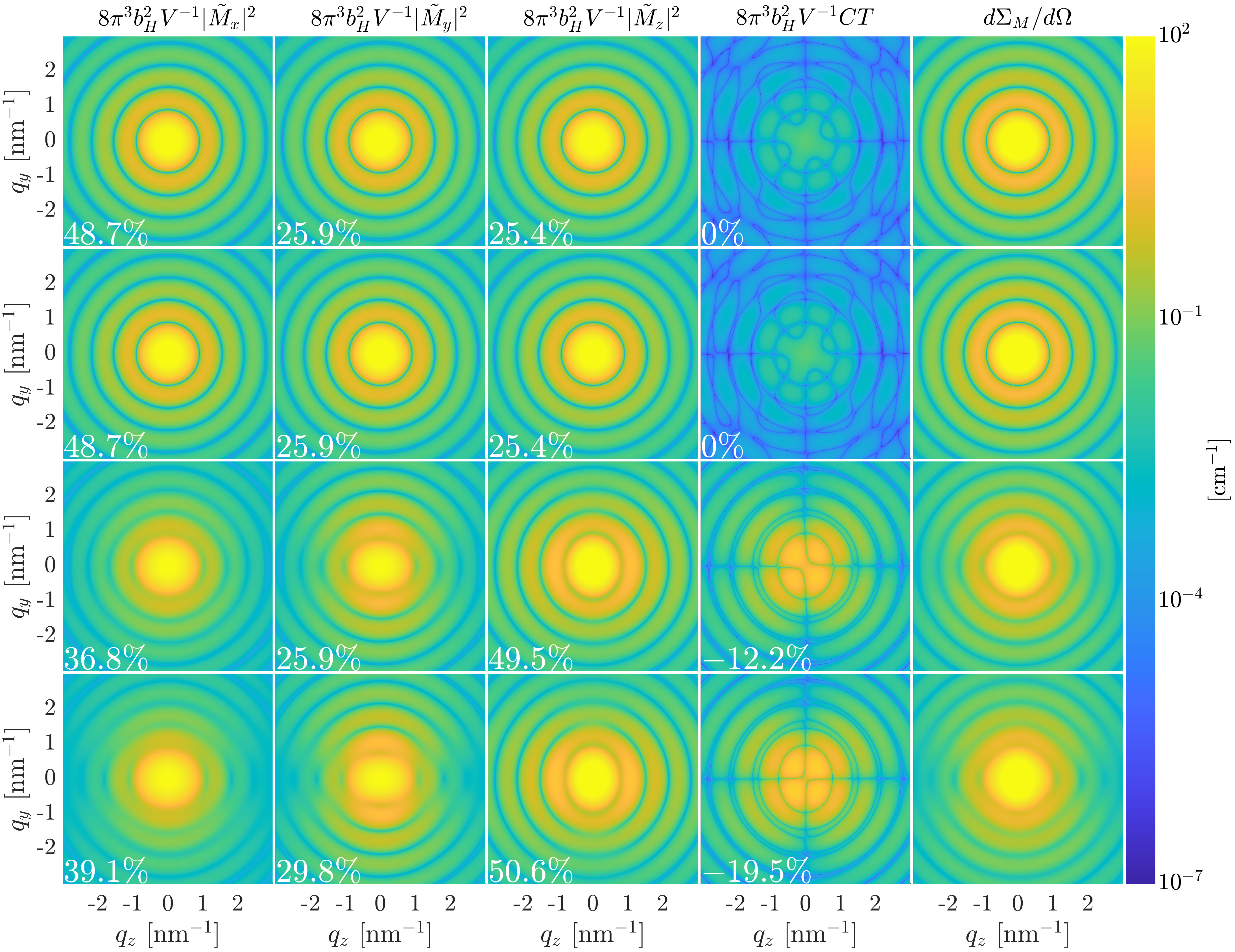}}
\caption{Decryption of the two-dimensional magnetic SANS cross section $d \Sigma_M / d \Omega$ in the remanent state ($B_0 = 0 \, \mathrm{T}$) into the individual magnetization Fourier components $|\widetilde{M}_x|^2$, $|\widetilde{M}_y|^2$, $|\widetilde{M}_z|^2$, and $CT = - (\widetilde{M}_y \widetilde{M}_z^{\ast} + \widetilde{M}_y^{\ast} \widetilde{M}_z)$ (see insets) (logarithmic color scale). Note that the respective Fourier components are multiplied with the constant $8 \pi^3 V^{-1} b_H^2$ (in order to have the same unit as $d \Sigma_M / d \Omega$), but not with the trigonometric functions in the expression for $d \Sigma_M / d \Omega$ [see equation~(\ref{eq:equation1})]. The $\%$~values specify the fraction of the respective Fourier component of the total $d \Sigma_M / d \Omega$ [see equation~(\ref{ratios}) and related text in the main paper]. The $CT$ (and hence the corresponding $\eta_\alpha$) can take on negative values, however, in this figure we show (due to the chosen logarithmic color scale) the absolute value of the $CT$. The data correspond to an ensemble of randomly-oriented $10 \, \mathrm{nm}$-sized nanomagnets. (upper row)~$K_s = 0$; (2nd row)~$K_s = 5.22 \times 10^{-23} \, \mathrm{J/atom}$; (3rd row)~$K_s = 26.1 \times 10^{-23} \, \mathrm{J/atom}$; (lower row)~$K_s = 52.2 \times 10^{-23} \, \mathrm{J/atom}$.}
\label{fig5}
\end{figure*}

\begin{figure*}[tb!]
\centering
\resizebox{1.0\columnwidth}{!}{\includegraphics{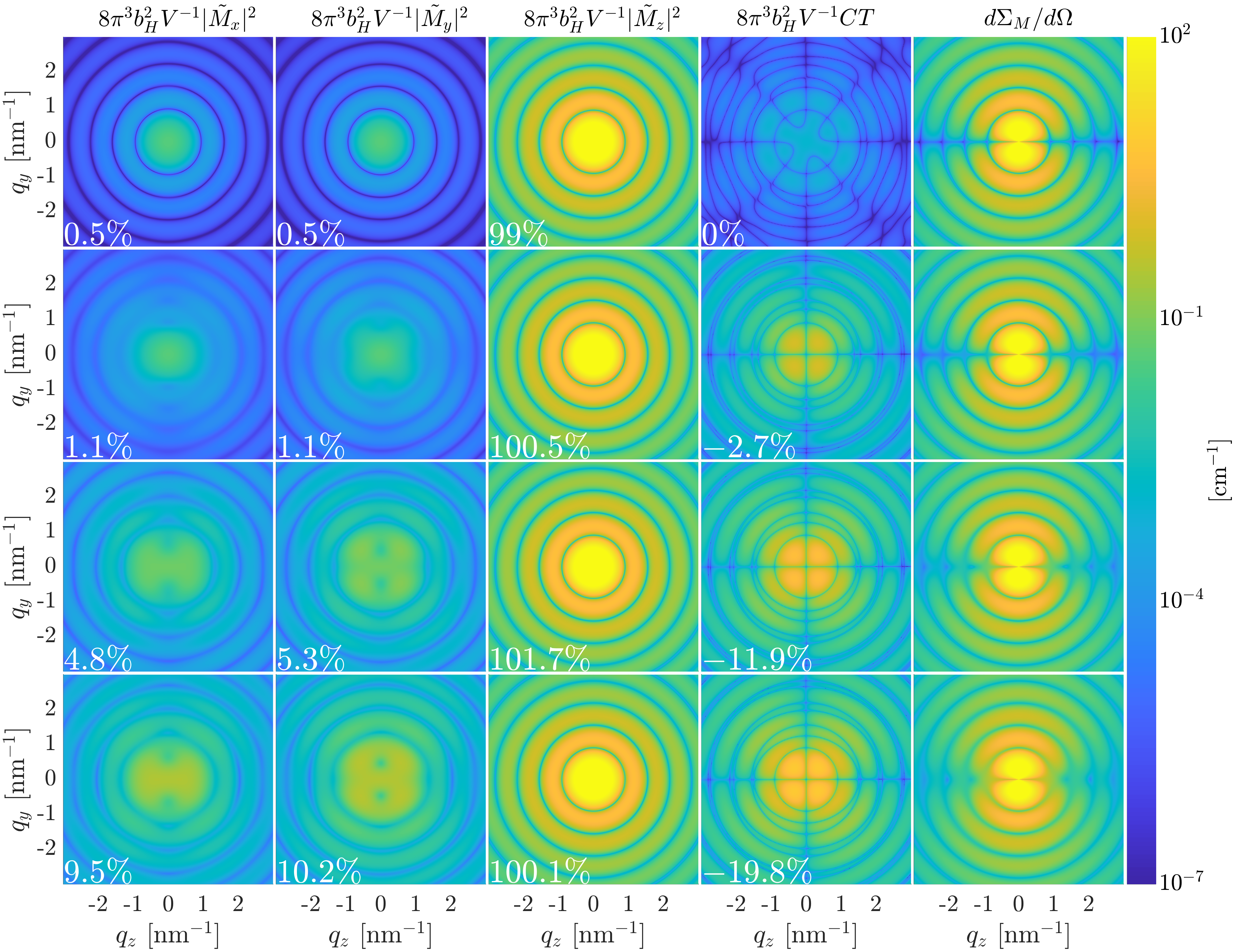}}
\caption{Same as Fig.~\ref{fig5}, but for $B_0 = 10 \, \mathrm{T}$.}
\label{fig6}
\end{figure*}

Fig.~\ref{fig5} displays the two-dimensional magnetic SANS cross section $d \Sigma_M / d \Omega$ of an ensemble of $10$-nm-sized nanomagnets in the remanent magnetization state along with the individual Fourier components $|\widetilde{M}_x|^2$, $|\widetilde{M}_y|^2$, $|\widetilde{M}_z|^2$, and the cross term $CT = - (\widetilde{M}_y \widetilde{M}_z^{\ast} + \widetilde{M}_y^{\ast} \widetilde{M}_z)$ [see equation~(\ref{eq:equation1})]. Fig.~\ref{fig6} shows the corresponding plots at a (nearly) saturating field of $B_0 = 10 \, \mathrm{T}$. We emphasize that the depicted scalar functions represent projections of the corresponding three-dimensional quantities onto the $q_y$$-$$q_z$~detector plane at $q_x = 0$ (see Fig.~\ref{fig2}). The surface anisotropy constant $K_s$ increases from the top to the bottom row in Figs.~\ref{fig5} and \ref{fig6}. It is seen that, generally, all the Fourier components are contributing to $d \Sigma_M / d \Omega$.

Near saturation (Fig.~\ref{fig6}), $d \Sigma_M / d \Omega$ is dominated for all values of $K_s$ by the isotropic ($\theta$~independent) $|\widetilde{M}_z|^2$~Fourier component and exhibits the characteristic $\sin^2\theta$~anisotropy with two maxima along the vertical direction [compare equation~(\ref{eq:equation1})]. Increasing $K_s$ enhances the contributions of both transversal Fourier components $|\widetilde{M}_x|^2$ and $|\widetilde{M}_y|^2$ and of the $CT$. Moreover, the latter contributions develop a pronounced angular anisotropy with increasing $K_s$.

At remanence (Fig.~\ref{fig5}), $d \Sigma_M / d \Omega$ and all the Fourier components are isotropic for small values of $K_s$ and become progressively more anisotropic with increasing $K_s$. For instance, $|\widetilde{M}_z|^2$ is initially isotropic and develops a pronounced angular anisotropy that is elongated along the $q_y$~direction for larger $K_s$. The $CT$ also develops an anisotropy with increasing $K_s$ with maxima roughly along the detector diagonals. An anisotropic magnetic SANS cross section at zero applied magnetic field of an ensemble of randomly-oriented nanoparticles has also been found in the micromagnetic continuum simulations of Vivas~et~al.~\cite{laura2020}. These authors did not consider the N\'{e}el surface anisotropy but included the magnetodipolar interaction. 

To quantify the fraction of the individual Fourier components in equation~(\ref{eq:equation1}) to the total magnetic SANS cross section $d\Sigma_M / d\Omega$, we have computed the following dimensionless quantity:
\begin{eqnarray}
\label{ratios}
\eta_{\alpha} =   \frac{1}{\pi q_{\mathrm{max}}^2} \int_{0}^{2\pi} \int_{0}^{q_{\mathrm{max}}} \frac{\alpha(q, \theta)}{d\Sigma_M/d\Omega} q dq d\theta ,
\end{eqnarray}
where $\alpha(q, \theta)$ is, respectively, given by $K |\widetilde{M}_x|^2$, $K |\widetilde{M}_y|^2 \cos^2\theta$, $K |\widetilde{M}_z|^2 \sin^2\theta$, and $K \, CT \sin\theta \cos\theta$ with $K = 8 \pi^3 b_H^2 V^{-1} $; $q_{\mathrm{max}}$ has been taken as $q_{\mathrm{max}} = 10 \, \mathrm{nm}^{-1}$. The sum of the $\eta_\alpha$ is then equal to one. The corresponding numbers are specified in the insets of Fig.~\ref{fig5}, and we note that the contribution related to $K \, CT \sin\theta \cos\theta$ can be negative, in contrast to the other three contributions which are strictly positive. Moreover, using the inequality $|\widetilde{M}_y \cos\theta - \widetilde{M}_z \sin\theta|^2 \geq 0$, it can easily be shown that the contribution $K \, CT \sin\theta \cos\theta$ is, however, always smaller than the sum of the other terms (as it must be). We emphasize that the color-coded plots in Fig.~\ref{fig5} show the respective Fourier components without the trigonometric functions in equation~(\ref{eq:equation1}), whereas the quantities $\eta_{\alpha}$ do contain the trigonometric terms. For $K_s=0$ and zero field, the contributions of $|\widetilde{M}_x|^2$, $|\widetilde{M}_y|^2$, and $|\widetilde{M}_z|^2$ to $d \Sigma_M / d \Omega$ are approximately equal (while $CT = 0$). This can be understood by noting the isotropy of these functions and by taking into account the trigonometric terms $\cos^2\theta$ (for $|\widetilde{M}_y|^2$) and $\sin^2\theta$ (for $|\widetilde{M}_z|^2$), which yield a factor of $1/2$ on azimuthal averaging [$\theta$~integration, compare equation~(\ref{ratios})].

\begin{figure*}[tb!]
\centering
\resizebox{1.0\columnwidth}{!}{\includegraphics{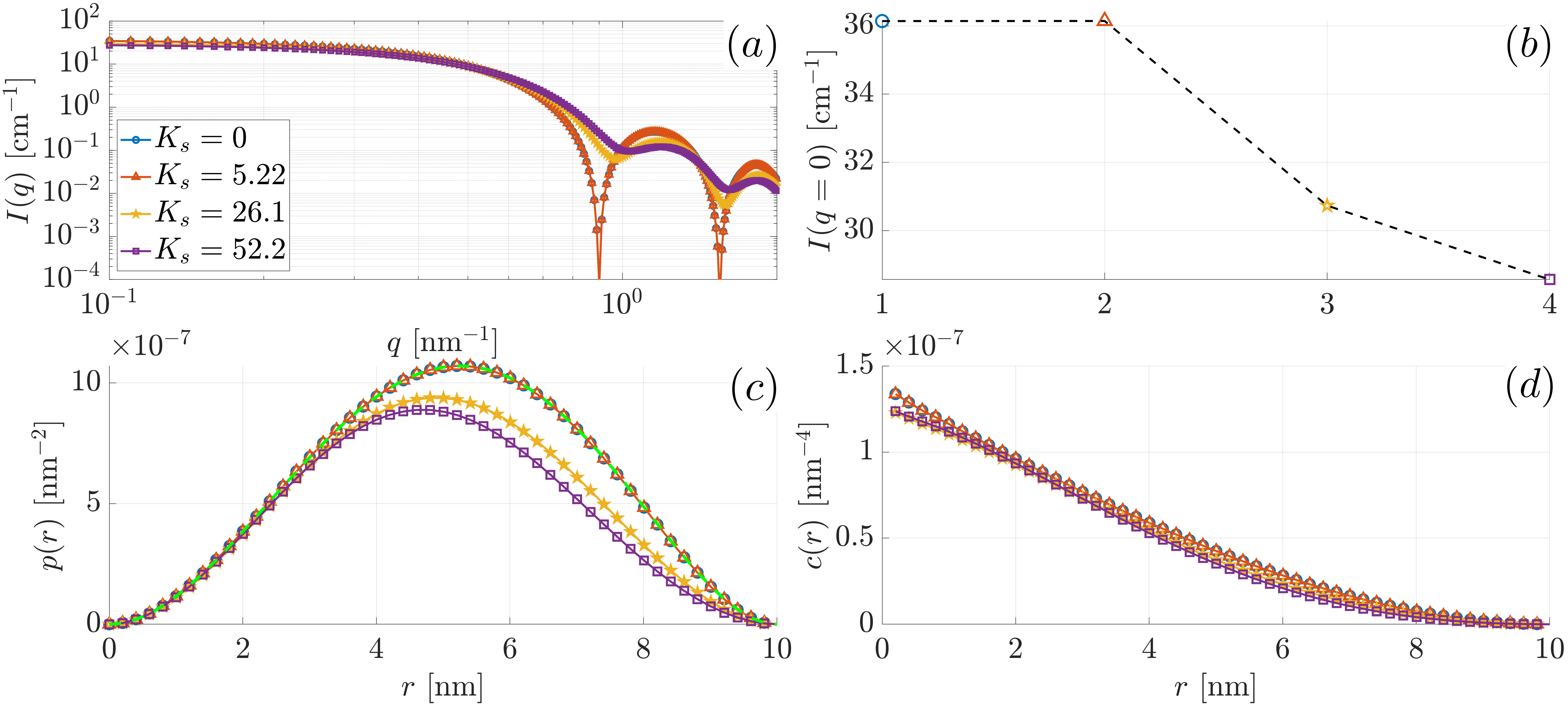}}
\caption{Effect of the surface anisotropy constant $K_s$ (in units of $10^{-23} \, \mathrm{J/atom}$, see inset) on (\textit{a})~the azimuthally-averaged magnetic SANS cross section $I(q) = \frac{d \Sigma_M}{d \Omega}(q)$ (log-log scale), (\textit{b}) the value of the magnetic SANS cross section at the origin, $I(q=0)$ versus $K_s$, (\textit{c})~the pair-distance distribution function $p(r)$, and (\textit{d}) the correlation function $c(r)$. The data correspond to the remanent state ($B_0 = 0 \, \mathrm{T}$), and the nanomagnets' diameter is $10 \, \mathrm{nm}$. The green dashed line in (\textit{c}) displays the analytical pair-distance distribution function for the case of a uniformly magnetized spherical particle [proportional to equation~\eqref{pvonreq}], where the magnitude is normalized to the maximum value from the numerical simulation in the case $K_s=0$.}
\label{fig7}
\end{figure*}

\begin{figure*}[tb!]
\centering
\resizebox{1.0\columnwidth}{!}{\includegraphics{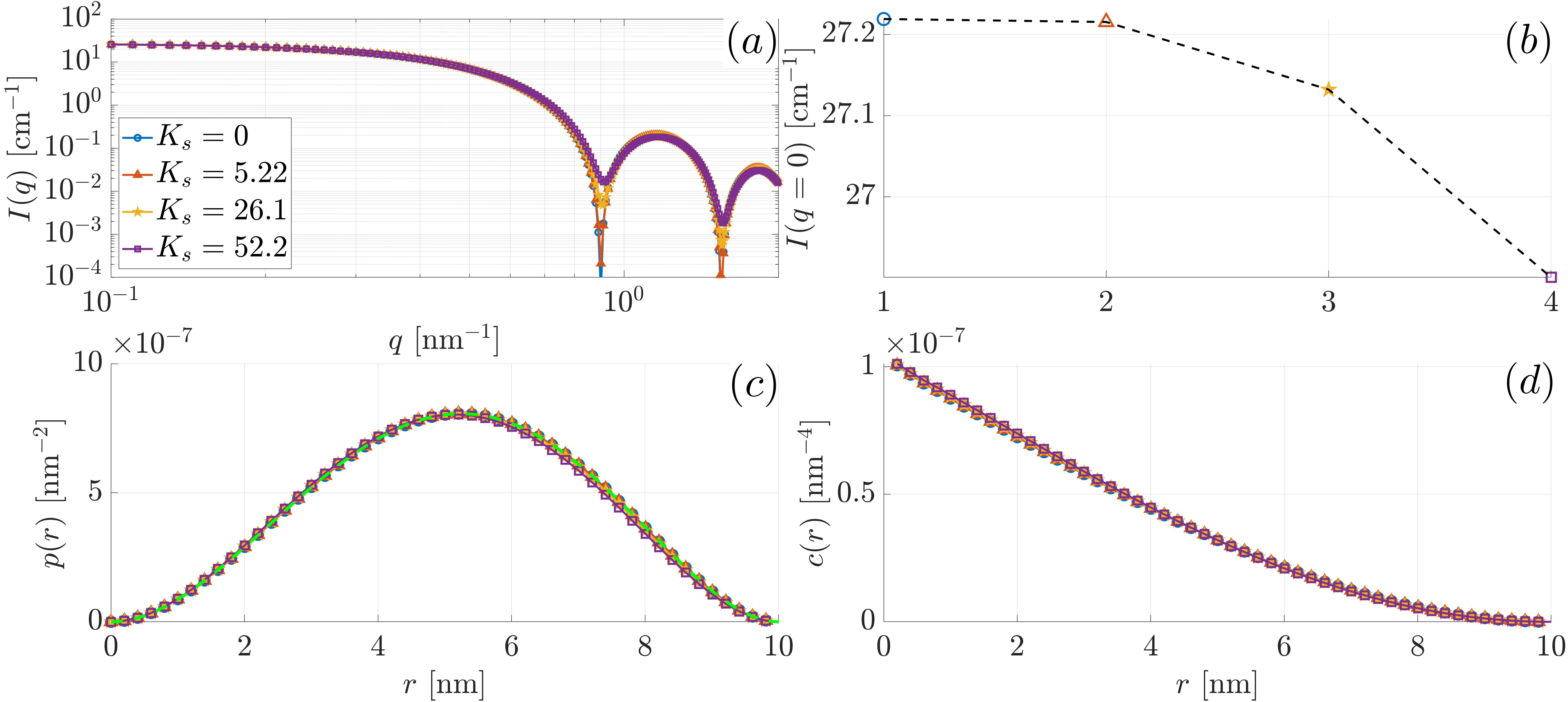}}
\caption{Same as Fig.~\ref{fig7}, but for $B_0 = 10 \, \mathrm{T}$.}
\label{fig8}
\end{figure*}

The effect of increasing $K_s$ on the $2\pi$~azimuthally-averaged $I(q) = d \Sigma_M / d \Omega$ and on $p(r)$ and $c(r)$ is shown in Fig.~\ref{fig7} for the remanent state and in Fig.~\ref{fig8} for $B_0 = 10 \, \mathrm{T}$. With increasing spin disorder (induced by an increasing $K_s$) we observe in Fig.~\ref{fig7}(\textit{a}) that (i)~the characteristic form-factor oscillations of $I(q)$ are progressively damped and (ii)~that the maxima in $I(q)$ shift to larger $q$~values. The smearing of the form-factor oscillations at increased $K_s$ mimicks the effect of a particle-size distribution function and/or of instrumental resolution. Therefore, in experimental situations, when such a data is fitted to \textit{e.g.}\ a sphere form factor with a distribution of particle sizes, an erroneous value for the particle size may result. At (quasi)saturation [Fig.~\ref{fig8}(\textit{c})] and for small $K_s$ at remanence [Fig.~\ref{fig7}(\textit{c})], we recover the analytically-known expressions for $I(q)$, $p(r)$, and $c(r)$ for uniformly magnetized spherical particles [equations~(\ref{homomagsans2}) and (\ref{pvonreq}), compare Figs.~\ref{fig7}(\textit{c}) and \ref{fig8}(\textit{c})]. We have also plotted in Figs.~\ref{fig7}(\textit{b}) and \ref{fig8}(\textit{b}) the $K_s$~dependence of the $q=0$~extrapolated value of $I(q)$. The quantity $I(q=0)$ is directly proportional to the static susceptibility $\chi(q=0)$ (as it can be measured with a magnetometer), which itself is proportional to the mean-square fluctuation of the magnetization per atom~\cite{lowde68}. We see that, as expected, the increase in $K_s$ has a large effect on $\chi(0)$, whereas the reduction is relatively small at $10 \, \mathrm{T}$.

\subsection{Effect of a particle-size distribution}
\label{sec4p2}

\begin{figure*}[tb!]
\centering
\resizebox{1.0\columnwidth}{!}{\includegraphics{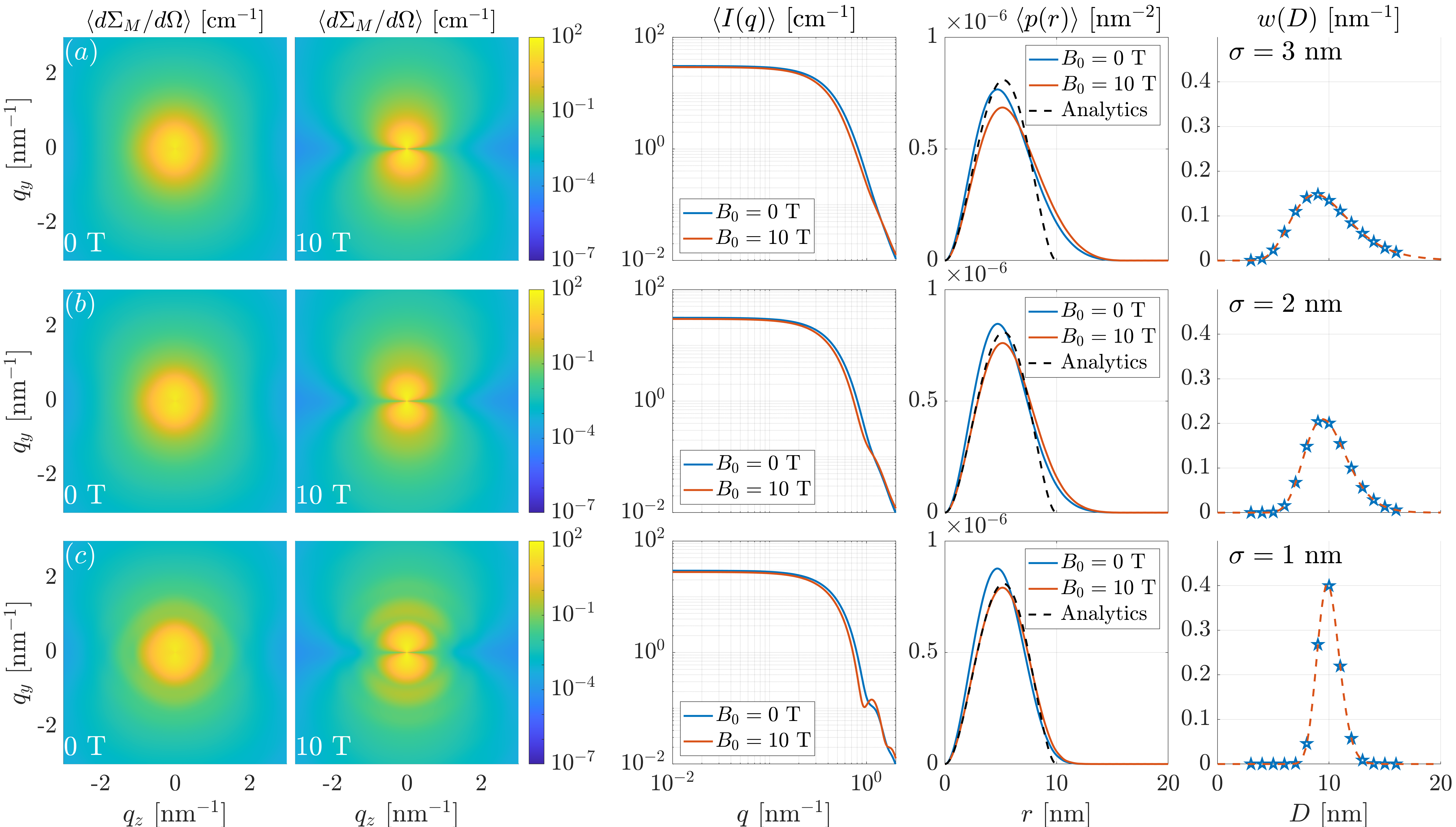}}
\caption{Effect of a lognormal particle-size distribution function on the SANS observables ($K_s = 52.2 \times 10^{-23} \, \mathrm{J/atom}$). Shown are the two-dimensional $\left \langle d\Sigma_M / d\Omega \right \rangle$, the corresponding azimuthally-averaged $\left \langle I(q) \right \rangle$, the pair-distance distribution functions $\left \langle p(r) \right \rangle$, and the particle-size distributions $w(D)$ for $\sigma$~values of (\textit{a}) $\sigma = 3 \, \mathrm{nm}$, (\textit{b}) $\sigma = 2 \, \mathrm{nm}$, and (\textit{c}) $\sigma = 1 \, \mathrm{nm}$. The nanoparticles' mean diameter (expectation value) was chosen as $\mu = 10 \, \mathrm{nm}$ in each case. The data correspond to the remanent ($B_0 = 0 \, \mathrm{T}$) and saturated ($B_0 = 10 \, \mathrm{T}$) magnetization state. The discrete particle-size classes are defined by the particle diameters $D = 3 \dots 16 \, \mathrm{nm}$ with an equidistant step size of $\Delta D = 1 \, \mathrm{nm}$. The black dashed $\left \langle p(r) \right \rangle$~curves are the analytically known solution for uniformly magnetized spheres of size $\mu = 10 \, \mathrm{nm}$ in the fully saturated state.}
\label{fig9}
\end{figure*}

In SANS experiments on nanoparticles one always has to deal with a distribution of particle sizes and shapes. The size of a particle has an important effect on its spin structure, \textit{e.g.}\ smaller particles tend to be uniformly magnetized, whereas larger particles may exhibit inhomogeneous spin structures. It is therefore also of interest to study the influence of a distribution of particle sizes on the magnetic SANS observables [$d\Sigma_{\mathrm{M}} / d\Omega$, $p(r)$, $c(r)$]. This has been done using a lognormal probability distribution function, which is defined as~\cite{krill98}:
\begin{align}
w(D) &= \frac{1}{\sqrt{2\pi D^2  
\operatorname{ln}\left(1 + \frac{\sigma^2}{\mu^2}\right)} } \exp\left(- 
\frac{\mathrm{ln}^2\left(\frac{D}{\mu}\sqrt{1 + \frac{\sigma^2}{\mu^2}}\right)}{2\operatorname{ln}\left(1 + \frac{\sigma^2}{\mu^2}\right)}
\right),
\end{align}
where $\mu$ denotes the expectation value and $\sigma^2$ is the variance, such that:
\begin{align}
    \mu &= \int_{0}^{\infty} w(D) D \; d D > 0 ,
    \\
    \sigma^2 &= \int_{0}^{\infty} w(D) (D - \mu)^2 \; d D ,
\end{align}
where the corresponding median $\mu^{\ast}$ is determined by the following relation:
\begin{align}
     \int_{0}^{\mu^{\ast}} w(D) \; d D &= \frac{1}{2}, &\mu^{\ast} &= \frac{\mu^2}{\sqrt{\mu^2 + \sigma^2}} .
\end{align}
For given values of $\mu$ and $\sigma$, the averaged magnetic SANS cross section $\langle ... \rangle$ is computed as:
\begin{eqnarray}
\left \langle \frac{d\Sigma_M}{d\Omega} \right \rangle = \sum_{\ell = 1}^{L} \frac{d\Sigma_{M,\ell}}{d\Omega}P_\ell ,
\label{lognorfunc}
\end{eqnarray}
where $P_\ell$ denotes the probability related to the particle-size class $D_\ell = 2 R_{\ell}$ (diameter), which is computed as:
\begin{align}
P_{\ell} =  \int_{D_{\ell}-\Delta D/2}^{D_{\ell}+\Delta D/2} w(D)\; d D .
\end{align}
Fig.~\ref{fig9} summarizes the results obtained for the magnetic SANS cross section and correlation function. As expected, one observes a smearing of the SANS cross section with increasing width $\sigma$ of the distribution, which becomes particularly visible in the azimuthally-averaged $\left \langle I(q) \right \rangle$~curves via the suppression of the form-factor oscillations. The angular anisotropy of the SANS cross section in the remanent state, which can be seen as a characteristic signature of the N\'{e}el anisotropy (compare also the lower row in Fig.~\ref{fig5}), becomes less pronounced for large $\sigma$. With increasing field, the $\left \langle p(r) \right \rangle$ are reduced in magnitude and the maximum is shifted to larger sizes (due to the suppression of the internal spin disorder).

\section{Conclusions and Outlook}
\label{sec5}

We have studied the spin structure and magnetic neutron scattering signal of an ensemble of randomly-oriented spherical nanomagnets using the Landau-Lifshitz equation with particular focus on the N\'{e}el surface anisotropy. Taking into account the isotropic exchange interaction, an external magnetic field, a uniaxial magnetic core anisotropy, and the N\'{e}el surface anisotropy, we compute the magnetic small-angle neutron scattering cross section and the pair-distance distribution function from the obtained equilibrium spin structures. The numerical results are compared to the well-known analytical expressions for uniformly magnetized particles. Upon increasing internal spin disorder (increasing surface anisotropy $K_s$), the pair-distance distribution function (at remanence) exhibits a systematic shift of its maximum to smaller $r$~values, and the total magnetic SANS cross section develops a characteristic anisotropic scattering pattern. The strength of the simulation methodology is that the field evolution of the individual Fourier components and their contribution to the magnetic SANS signal can be monitored. Atomistic and micromagnetic continuum simulations have contributed and will continue to contribute to the fundamental understanding of magnetic SANS. In our future work, we will focus on the inclusion of both the intraparticle and the interparticle dipole-dipole energy and the Dzyaloshinskii-Moriya interaction, which will give rise to more complicated spin textures (\textit{e.g.}\ vortex-type structures), in particular for larger particle sizes. Moreover, it is of interest to compare the N\'{e}el anisotropy to other phenomenological expressions for the surface anisotropy, such as energy densities of the type $\pm \frac{1}{2} K_s (\mathbf{m} \cdot \mathbf{n})^2$, where $\mathbf{n}$ is the unit normal vector to the surface (instead of the $\mathbf{u}_{ij}$), or to the case of a truly random surface anisotropy, where the $\mathbf{u}_{ij}$ are random vectors. In this regard, the present first atomistic simulations may be considered as the starting point towards a more complete description of magnetic SANS.

\acknowledgments{Michael Adams and Andreas Michels thank the National Research Fund of Luxembourg for financial support (AFR Grant No.~15639149).}


%

\end{document}